\documentclass{appolb}
\newcommand{\beq}{\begin{equation}}
\newcommand{\eeq}{\end{equation}}
\newcommand{\beqa}{\begin{eqnarray}}
\newcommand{\eeqa}{\end{eqnarray}}
\newcommand{\beql}[1]{\begin{eqletters}\label{#1}\begin{eqnarray}}
\newcommand{\eeql}{\end{eqnarray}\end{eqletters}}

\newcommand{\erdos}{Erd\H{o}s}
\newcommand{\renyi}{R\'enyi}

\renewcommand{\)}{\right)}
\renewcommand{\c}{\mathbf{c}}

\newcommand{\m}{\mathbf{m}}
\newcommand{\binom}[2]{\left( \begin{array}{c} #1\\#2 \end{array} \right)}
\newcommand{\T}{\mathbf{T}}
\newcommand{\J}{\mathbf{J}}
\newcommand{\E}{\mathbf{E}}
\newcommand{\1}{\mathbf{1}}
\newcommand{\z}{\mathbf{z}}
\newcommand{\x}{\mathbf{x}}
\newcommand{\g}{\mathbf{g}}
\renewcommand{\r}{\mathbf{r}}
\newcommand{\h}{\mathbf{h}}
\renewcommand{\e}{\mathbf{e}}
\newcommand{\ave}[1]{\left\langle #1 \right\rangle}
\usepackage{graphicx}
\begin{document}
%
\title{Random Graph Models with Hidden Color \thanks{
Presented at the Workshop on Random Geometry in Krakow, May 2003}}
\author{Bo S\"oderberg \address{Complex Systems Div., Dept. for
 Theoretical Physics, Lund University, S\"olvegatan 14 A, S-223 62
 LUND, Sweden}}
\maketitle
\begin{flushright}
LU TP 03-34\\
Submitted to Acta Physica Polonica B
\end{flushright}
\begin{abstract}
We demonstrate how to generalize two of the most well-known random
graph models, the {\em classic random graph}, and {\em random graphs
with a given degree distribution}, by the introduction of hidden
variables in the form of extra degrees of freedom, color, applied to
vertices or stubs (half-edges). The color is assumed unobservable, but
is allowed to affect edge probabilities. This serves as a convenient
method to define very general classes of models within a common
unifying formalism, and allowing for a non-trivial edge correlation
structure.
\end{abstract}

\section{Introduction}

The availability of data on real-world networks, \eg from information
technology and molecular biology, has seen a dramatic increase in the
last decades.  This has led to a correspondingly increased interest in
the theoretical modelling of networks.

Typically the growth of a real-world network is not entirely
deterministic but contains stochastic elements, and statistical models
are required that are conveniently formulated in terms of {\em
ensembles} of graphs.  Typical real-world networks are not static but
change with time, and much of the focus has been on {\em dynamical}
models, where one attempts to describe the growth and evolution of a
network.

Here we will focus on {\em static} random graph models, describing a
snapshot of a network in terms of a fixed ensemble of graphs, without
regard to how the network was formed.  By a {\em random graph} we will
mean a member of such an ensemble. In particular, we will be mostly
interested in {\em sparse} random graphs, where the typical vertex degree
does not grow with the size of the graph.

There is a vast spectrum of such models around. Some of these are not
entirely random, in the sense that they are based on an underlying
regular network -- i.e. a lattice -- which is then modified in a
random fashion.

Our focus will be on {\em purely random} graphs, where such an
underlying regularity is absent. A number of more or less unrelated
models of this type have been investigated, and it would obviously be
desirable to devise a unified description in terms of a {\em general
class of ensembles}, where more specialized models appear as special
cases of one and the same general formalism, while maintaining the
computability of local and global graph characteristics of interest,
such as degree distributions, small subgraph abundancies, component
size distributions, and global connectivity properties.

The most well-known purely random model is the {\em classic Random
Graph} of {\erdos} and {\renyi} \cite{ErRe60}, to be referred to as
{\em RG}. In its sparse version it is defined as follows. For a given
set of $N$ nodes, every pair of nodes is connected by an edge
independently with probability $p=c/N$ in terms of a given parameter
$c$ that asymptotically defines the average degree. This model has
many interesting properties, such as an asymptotically Poissonian
degree distribution, and a phase transition at $c=1$, above which a
giant connected component is formed.  However, it fails to describe
most real-world networks.

A more general model that has been much studied is {\em Random Graphs
with a given Degree Distribution} \cite{BenCan,Lucz92,MoRe98,Newm01},
or Degree-driven Random Graphs ({\em DRG}), where an asymptotic degree
distribution is given, suitably transformed into a definite degree
sequence for a given graph size.  In terms of this a random graph is
defined as a uniformly random member drawn from the set of graphs
having the given degree sequence, possibly subject to additional
constraints (\eg by demanding the graph to be simple, \ie
non-degenerate).  DRG models suffer from an intrinsic lack of edge
correlations, atypical of real-world networks; as a result, they are
often referred to as uncorrelated random graphs.

In a sequence of papers \cite{Sod02,CDRG,CDRG2}, I have explored the
use of hidden coloring, either of vertices or of stubs, to define more
general random graph models. The resulting models can be seen as
colored extensions of RG and DRG.  As will be shown in this paper, the
hidden color provides a convenient means for defining very general
classes of random graph models, where much of the limitations of the
uncolored models can be done away with, while the computability of
interesting properties is maintained.

The considered classes of models will be compared with respect to a
few basic properties: The degree distribution, the abundancy of
arbitrarily given small subgraphs, the size distribution of connected
components, and the phase transition where a giant component appears.

The plan of the paper is as follows.
In section 2, we will discuss a few fundamental concepts needed in the
subsequent sections. In section 3, we will review the definitions of
the models to be considered. A comparative analysis of a selected set
of characteristics, as derived in the different model classes, is
presented in section 4. Section 5, finally, contains a concluding
discussion.

\section{Basic Concepts and Methods}

All models to be considered in this paper will be of sparse random
graphs, where the degrees (connectivities) of vertices stay finite as
the size $N$ of the graph grows to infinity.
In particular, this means that the total number of edges will scale as
$N$, and that the probability of a connection between an arbitrary
pair of nodes will scale as $1/N$.

A simple local characteristic of a graph ensemble is its {\em degree
distribution}, $\{p_m\}$. This is often conveniently described in
terms of its {\em generating function},
\beq
\label{H}
	H(x) \equiv \sum_m p_m x^m.
\eeq
It obviously satisfies $H(1)=1$, and yields upon repeated
differentiation at $x=1$ successive {\em combinatorial moments} of the
degree,
\beqa
\label{H-mom}
	H'(1) = \ave{m},\;
	H''(1) = \ave{m(m-1)},\;
	H'''(1) = \ave{m(m-1)(m-2)},
\eeqa
etc., while the individual $p_m$ can be obtained by repeated
differentiation at $x=0$.

Generating functions of this type are convenient when analyzing a
probability distribution $P_k$ of an integer variable $k$ that is the
{\em sum} of several independent contributions, $k=\sum_i k_i$, in
which case the generating function $f(z)=\sum_k P_k z^k$ for the
distribution of $k$ is simply the {\em product} of the corresponding
generating functions for the distribution of each contribution.

\section{The Models}

Here follows a brief introduction to the models to be considered.

\subsection{The Classic Model -- RG}

The classic random graph (RG) \cite{ErRe60} has been thoroughly
analyzed over the years \cite{Boll01,Jans00}.  It is a model of simple
(non-degenerate) labelled graphs with a given set of $N$ nodes,
although it can be easily extended to include also non-simple graphs
\cite{FKP89}.  We will consider its sparse version. It comes in two
essentially equivalent versions, one with a fixed number of edges, the
other with a fixed probability for each possible edge; we will stick
to the latter.

Sparse RG has a single real parameter $c>0$ controlling the abundance
of edges. For a given graph size $N$ and a given value of $c$, an
ensemble of graphs is defined as follows. Each of the $N(N-1)/2$ pairs
of distinct nodes independently is connected by an undirected edge
with a common probability $p=c/N$ (assuming $N>c)$.

\subsection{Inhomogeneous Random Graphs -- IRG}

The classic RG model as described above can by generalized in a
straightforward way by assigning color to vertices and allowing edge
probabilities to be color-sensitive; the resulting class of models
will be referred to as {\bf IRG}, for inhomogeneous random graphs
\cite{Sod02}.

A definite IRG model is specified in terms of
\begin{itemize}
\item a color space, taken as $[1,\dots,K]$;
\item a {\em color distribution} $\{r_a>0,\;a=1,\dots,K\}$, with
$\sum_a r_a = 1$;
\item a real, symmetric {\em color preference matrix} $\c=\{c_{ab} \ge
0\}$.
\end{itemize}
For a given graph size $N$, such a model is implemented as follows.
\begin{enumerate}
\item Assign to each node independently a random color $a$, drawn from
the given distribution $\{r_a\}$.
\item Connect each pair of distinct nodes independently with
probability $c_{ab}/N$, where $a$ and $b$ are the respective colors of
the two nodes.
\end{enumerate}
By considering the color as unobservable or hidden, the resulting
ensemble of colored graphs yields a specific ensemble of plain graphs,
distinct from an RG ensemble, as will be shown below.  The role of the
hidden color is to enable non-trivial edge correlations. IRG defines a
class of graph ensembles much more general than RG.

\subsection{Random graphs with a given degree distribution -- DRG}

The classic RG model is limited to a Poissonian degree distribution.
A more general class of models that has recently attracted the
attention of several workers in the statistical physics community is
random graphs with a given degree distribution
\cite{BenCan,MoRe98,Newm01}, to be referred to as {\em DRG} (for
degree-driven random graphs).\footnote{Variants of this approach have
been referred under various names, such as {\em equilibrium random
graphs} and {\em uncorrelated random graphs}.} This approach allows
for an arbitrary degree distribution.

There are two common variants of DRG. One is given by restricting the
ensemble to simple (non-degenerate) graphs, where self-couplings and
multiple connections are banned. In the other (the configuration
model) one allows for degenerate graphs. For ease of analysis, we will
focus on the latter version where degeneracies are allowed.

Some notation: A node with degree $m$ is considered to possess $m$
{\em stubs}, each of which defines a point of attachment for an edge
endpoint. The total number of stubs $M=\sum_i m_i$ in a graph
obviously must be even, being equal to the total number of edge
endpoints, i.e. twice the number of edges.

A specific DRG model is defined by specifying an arbitrary degree
distribution $\{p_m\}$.  For a fixed graph size
$N$,\footnote{We disregard impossible cases of an odd $N$ with a degree
distribution supporting only odd degrees.} the corresponding ensemble
is implemented as follows.
\begin{enumerate}
\item For each node, draw its degree independently from the given
distribution. Redo until the total sum of the degrees is even.
\item Define random edges by performing a completely random pairing
within the resulting even-numbered set of $M$ stubs.
\end{enumerate}
This leads to an ensemble of pseudographs, where degeneracies may
appear, in the form of self-connections (tadpoles) or multiple edges
between the same pair of nodes.

The random stub pairing is reminiscent of the combinatorics associated
with Gaussian integrals; indeed, a relation exists between DRG models
and certain miniature field theories \cite{BuCoKr01,DoMeSa02}.

\subsection{DRG plus color -- CDRG}

Also the DRG class of models can be generalized, by utilizing a
coloring of {\em stubs}, which turns out to be the most natural
choice, and then allowing the stub pairing to be color-sensitive
\cite{CDRG,CDRG2}.  The resulting very general class of models will be
referred to as CDRG, for Colored DRG.

With colored stubs, it is natural to consider the color-specific stub
content of a node, its {\em colored degree}. With $K$ colors to choose
between, the colored degree is conveniently represented by an integer
vector $\m=(m_1,\dots,m_K)$, with the individual elements $m_a$
counting the number of stubs with color $a$. Obviously the plain
degree $m$ is obtained by summing up the elements of the colored
degree, $m=\sum_a m_a = \1\cdot\m$, in terms of the uniform vector
$\1=(1,\dots,1)$.

Then it is also natural to consider the probability distribution of
such colored degrees, a {\em colored degree distribution}
$\{p_{\m}\}$. Such a distribution can be represented by a {\em
multivariate generating function}, $H(\x)=\sum_{\m} p_{\m} \x^{\m}$,
where $\x^{\m} = \prod_ax_a^{m_a}$, satisfying the normalizing
condition $H(\1)=1$. From this, {\em multivariate combinatorial
moments} can be derived by repeated differentiation at $\x=\1$,
e.g. $\partial_a H(\x=\1)=\ave{m_a}$, $\partial_a \partial_b
H(\x=\1)=\ave{m_am_b-m_a\delta{ab}}$, etc.
A specific CDRG model is defined by specifying
\begin{itemize}
\item A color space, taken as $[1,\dots,K]$;
\item A colored degree distribution $\{p_{\m}\}$;
\item A real, symmetric color preference matrix $\T=\{T_{ab} \ge 0\}$,
such that $\T\ave{\m}=\1$.
\end{itemize}
We will for simplicity assume that the colored degree distribution is
such that all moments are defined.

For a given graph size $N$, such a model is implemented as follows.
\begin{enumerate}
\item For each node, draw its colored degree independently from the
given distribution. Redo until the total sum of the (plain) degrees is
even.
\item Define random edges by performing a weighted random pairing
within the resulting even-numbered set of $M$ stubs, such that the
probability for each of the $(M-1)!!$ possible pairings has a
statistical weight proportional to the product over all edges of a
factor given by $T_{ab}$, where $a,b$ are the colors of the stubs it
connects.
\end{enumerate}
This class of models obviously collapses to DRG for the case of a
single color, in which case the matrix $\T$ collapses to a single
number, given by $1/\ave{m}$ by virtue of the constraint
$\T\ave{\m}=1$.

The constraint on $\T$ is convenient for the forthcoming analysis, and
ensures that the total number of $ab$-edges asymptotically approaches
the value $N\ave{m_a}T_{ab}\ave{m_b}$, which upon summing over $b$
yields the correct asymptotic number of $a$-stubs as $N\ave{m_a}$.

The combinatorics of the weighted random pairing yields the following
asymptotic results. The probability that two arbitrary stubs with
known colors $a,b$ will be paired with each other is $T_{ab}/N$.  This
implies that the the probability for two random nodes with respective
colored degrees $\m,\m'$ will be connected is
$\sum_{ab}m_aT_{ab}m_b'/N$, which reduces (as it must!) to $\ave{m}/N$
if the degrees are not known.

A less general class of models, similar in spirit to CDRG but
restricted to homogeneously colored vertices (i.e. with vertex
coloring rather than stub coloring), has also been investigated
\cite{Newm03}.

\section{Analysis}

Below follows a comparative analysis, where we review a selected set
of local and global characteristics for random graphs as drawn from
models of the different types, with a focus on the asymptotic limit $N
\to \infty$. For a more detailed analysis, we refer to the paper
\cite{CDRG2} and references therein.

\subsection{Asymptotic Degree Distributions}

First we will derive the resulting asymptotic degree distributions,
where not defined in the model specifications.

\subsubsection{RG degree distributions}

In a graph drawn from an RG model as described above, each node has
$N-1$ possible connections, each independently realized with
probability $c/N$. Thus, the degree $m$ of a random node will obey a
binomial distribution, $\binom{N-1}{m} (c/N)^m(1-c/N)^{N-1-m}$.  As
$N\to\infty$ with fixed $c$, this approaches an asymptotic
distribution $\{p_m\}$, given by a {\em Poissonian} with average $c$,
\beq
 	p_m = e^{-c}\frac{c^m}{m!},
\eeq
with the corresponding generating function $H(x)=e^{c(x-1)}$.

\subsubsection{IRG degree distributions}

Choose a random node in a large graph from an IRG ensemble. It has the
color $a$ with probability $r_a$. For large $N$ there are $\sim Nr_b$
other nodes with color $b$; each of these is connected to the chosen
node independently with probability $c_{ab}/N$. Thus, for a node of
given color $a$, we asymptotically expect its number of $b$-neighbors
to follow a Poissonian distribution with average $c_{ab} r_b$, and its
total degree to follow a Poissonian with average $C_a \equiv \sum_b
c_{ab} r_b$.  Averaging over $a$ yields the asymptotic degree
distribution
\beq
\label{pm_irg}
	p_m = \sum_a r_a e^{-C_a} \frac{C_a^m}{m!},
\eeq
with the generating function
\beq
	H(x) = \sum_a r_a e^{C_a(x-1)},
\eeq
which describes a {\em Poissonian mix}. This implies the following
convexity constraint on the possible degree distributions:
\beq
	p_m^2 \le \frac{m+1}{m} p_{m-1} p_{m+1},
\eeq
for $m>0$. Conversely, any degree distribution obeying this constraint
can, at least in principle, be realized with a suitable IRG model,
possibly with infinitely many colors.

\subsubsection{DRG degree distributions}

The degree distribution is considered given in a DRG model, and so is
in principle free to choose. For ease of analysis, we shall restrict
our considerations to cases where all moments $\ave{m^n}$ exist,
barring power-behaved distributions, which otherwise are interesting
in their own right.

\subsubsection{CDRG degree distributions}

In a CDRG model, a colored degree distribution is given, from which
the plain degree distribution can be extracted directly. Its
generating function $H(x)$ is obtained simply by evaluating the
multivariate generating function (with the same name) for the colored
degree distribution with a homogeneous argument, $H(x) = H(x\1) \equiv
H(x,\dots,x)$.

Since the colored degree distribution is free to choose, so is the
plain one, and there are obviously many CDRG models with a given
degree distribution.

\subsection{Small Subgraph Statistics}

The combinatorial moments of the degree distribution are simply
related to the expected numbers of subgraphs in the form av stars.
More general local characteristics can be expressed in terms of the
number of copies of an arbitrary small graph $\gamma$ found as
subgraphs of a large random graph $G$. We will be interested in the
expected number of copies in the asymptotic limit $N\to\infty$.

The clustering properties of a graph are often analyzed in
terms of the probability of two neighbors of a node to be connected;
this is seen to be related to the number of simple triangles, i.e. the
number of subgraphs $\gamma$ in the form of a mutually connected
triple of nodes.

Thus, assume an arbitrary small connected graph $\gamma$ to be given,
having $v\ll N$ vertices and $e\ll N$ edges. We can estimate its
expected number $\ave{n_{\gamma}}$ of distinct occurrencies as a
subgraph in a random graph $G$ of size $N$ as follows.

A particular possible embedding of $\gamma$ in $G$ is defined by
mapping the ordered set of $v$ nodes in $\gamma$ onto a target set
given by an ordered $v$-subset of the $N$ vertices in $G$. There are
$N!/(N-v)! \approx N^v$ such sets.  However, for a target set to
define a valid subgraph position, each edge in $\gamma$ must be mapped
onto an existing edge in the target set.

For the models considered, the expected count $\ave{n_{\gamma}}$ can
be derived from Feynman-like rules, with model-specific vertex and
node factors as well as the usual symmetry factors.

\subsubsection{Small subgraphs in RG}

The RG model describes simple random graphs, which can have only simple
subgraphs.  For each of the $\sim N^v$ possible embeddings of a simple
$\gamma$, the probability for the corresponding set of $e$ target
edges to exist is $(c/N)^e$.

Thus, naively, the expected number of occurrencies should be
$N^{v-e}c^e$. If $\gamma$ has a non-trivial isomorphism group, i.e. a
symmetry under some permutation of its vertices, the naive result has
to be divided by the order $S_{\gamma}$ of the symmetry group. This
leaves us with the following simple rules for the asymptotically
expected subgraph count $\ave{n_{\gamma}}$.
\begin{itemize}
\item For each node in $\gamma$, associate a factor $N$.
\item For each edge in $\gamma$, associate a factor $c/N$.
\item Multiply the node and edge factors, and divide the result by the
symmetry factor $S_{\gamma}$.
\end{itemize}
Since $\gamma$ is assumed connected, we have $e\ge v-1$, and $e-v+1$
counts its number of loops. Thus, the expected count scales as $O(N)$
for a {\em tree}, and as $O(1)$ for a one-loop $\gamma$, while it
vanishes asymptotically for $\gamma$ with several loops. This is
typical of a sparse random graph -- loops are scarce.

As an example illustrating the lack of correlations in an RG ensemble,
consider subgraphs in the form of a $v$-chain, i.e. a set of $v$ nodes
connected in an open chain; the expected counts show a simple
geometric behaviour, $\ave{n_{\gamma}} = N c^{v-1}/2$.

\subsubsection{Small subgraphs in IRG}

Also in IRG, graphs are simple, so also here, we must assume $\gamma$
to be simple.  Generalizing the arguments used for RG, we get the
following rules for the asymptotically expected number
$\ave{n_{\gamma}}$.
\begin{itemize}
\item Associate with each node in $\gamma$ an independent color $a$,
and a corresponding factor $Nr_a$.
\item Associate with each edge in $\gamma$ a factor $c_{ab}/N$, where
$a,b$ are the node colors at its endpoints.
\item Multiply all node and edge factors, sum over the node colors,
and divide the result by the symmetry factor $S_{\gamma}$.
\end{itemize}
Again, expected counts for tree subgraphs scale as $O(N)$, and those
for connected one-loop subgraphs as $O(1)$.

Non-trivial edge correlations are possible in IRG, as illustrated by
$v$-chain subgraphs, where the hidden color in an IRG ensemble enables
the expected counts to deviate from the simple geometric behavior
found for a plain RG ensemble; instead it takes the form of a mix of
geometric sequences.

\subsubsection{Small subgraphs in DRG}

Since a DRG ensemble of the kind we are considering allows for
degeneracies, we will have to consider also possibly degenerate
subgraphs, with loops of length one or two. Since subgraphs with loops
are suppressed due to the sparsity, just as in RG and IRG,
degeneracies will turn out not to be very important.

The expected number of copies of $\gamma$ with a fixed set of target
nodes can be calulated as follows. Consider a node in the target set
with actual degree $m$, that defines the target for a node with degree
$k$ in $\gamma$. The corresponding $k$ target edges can be chosen
among the $m$ existing ones in $m_k\equiv m!/(m-k)!$ distinct ways.
This can be shown to yield the following rules for calculating the
asymptotic $\ave{n_{\gamma}}$ for a DRG model.
\begin{itemize}
\item Associate with each node with $k$ stubs in $\gamma$ a factor $N
\ave{m_k}$.
\item Associate with each edge in $\gamma$ a factor $1/(N\ave{m})$.
\item Multiply the node and edge factors, and divide the result by the
symmetry factor $S_{\gamma}$, including possible contributions from
edge permutations and flips for the case of a non-simple $\gamma$.
\end{itemize}
Here, $\ave{m_k}$ stands for the $k$th combinatorial moment, defined
by $\partial_z^k H(z=1)=\ave{m (m-1) \dots (m-k+1)}$ (see
eq. (\ref{H-mom})).

For a $v$-chain, the expected count becomes $N \ave{m}^{3-v}
\ave{m(m-1)}^{v-2} / 2$, displaying simple geometric behavior just as
for the case of RG, illustrating the lack of edge correlations in DRG.

\subsubsection{Small subgraphs in CDRG}

To analyze the subgraph statistics for a CDRG model, we will need the
colored generalizations of the combinatorial moments,
\beq
\label{MCMom}
	E_{abc\dots} \equiv \partial_a \partial_b \partial_c \dots H(\z=\1),
\eeq
with the lowest ones given by $E_a\equiv\ave{m_a}$,
$E_{ab}\equiv\ave{m_am_b-m_a\delta_{ab}}$, etc.
Generalizing the argument used for DRG, one can derive the following
rules \cite{CDRG2} for the asymptotically expected subgraph counts in
a CDRG model.
\begin{itemize}
\item Associate with each stub in $\gamma$ an independent color label.
\item Associate with each node in $\gamma$ having $k$ stubs a factor
$N E_{abc\dots}$, where $a,b,c\dots$ are the $k$ associated color
labels.
\item Associate with each edge in $\gamma$ a factor $T_{ab}/N$, where
$a,b$ are the color labels associated with the stubs at its endpoints.
\item Multiply the node and edge factors, sum over the associated
colors, and divide the result by the symmetry factor $S_{\gamma}$,
including possible contributions from edge permutations and flips for
the case of a non-simple $\gamma$.
\end{itemize}
Note how these reduce to the DRG rules for the case of a single color.

For the case of a $v$-chain, the expected count becomes a mix of
geometric sequences, showing how the coloring also for CDRG enables
non-trivial edge correlations, just as was the case for IRG.

\subsection{Connected Component Sizes}

Next we turn to an analysis of the global connectivity characteristics
of a random graph. These are simplest described in terms of the sizes
of the connected components of the graph.

Thus, we will be interested in the size distribution $P_n$ of a
connected component of a random graph, as revealed from a randomly
chosen initial node by recursively exploring edges leading to new
nodes until the entire component is revealed. For sparse random graphs
in the asymptotic limit $N\to\infty$, any finite component is almost
surely a tree, since any extra connections will be suppressed by
factors of $1/N$. Thus, the revelation of such a component can be
described as a {\em branching process}, with properties depending on
the specific model considered.

The asymptotic component size distribution is conveniently analyzed in
terms of its generating function,
\beq
	g(z) \equiv \sum_n P_n z^n.
\eeq
For the models considered, $g(z)$ or a set of related functions will
satisfy recursive equations that determine the sought distribution.

\subsubsection{Component sizes in RG}

For an RG model, the component size distribution can be estimated as
follows, as long as the component remains small. For each revealed
node $i$, the number $k$ of branches to new nodes obeys a Poissonian
distribution asymptotically, $p_k \sim e^{-c}c^k/k!$, since there are
$\sim N$ remaining unrevealed nodes, each of which connects to $i$
with probability $c/N$.

This yields the recursive equation $g(z) = z e^{-c} \sum_k c^k g(z)^k
/ k!$, to be understood as follows. The initial factor of $z$ accounts
for the initial node, while each term in the sum describes the case
where it has a distinct number $k$ of neighbors. The factor $e^{-c}
c^k / k!$ represents the probability for this case, and the factor
$g(z)^k$ encodes the fact that each of the $k$ neighbors defines a
subtree statistically identical to the full tree.

The recursion can be simplified to read
\beq
\label{g_RG}
	g(z) = z e^{c(g(z)-1)},
\eeq
which should be interpreted as an {\em iterated map} for the value of
$g$ for a given value of $z$, a {\em stable fixed point} of which
defines the physical value.

As a curiosity, eq. (\ref{g_RG}) can be written as $F(cg(z)) = z
F(c)$, with $F(c)=ce^{-c}$, with the explicit solution $g(z) =
F^{-1}(zF(c))/c$, with the inverse of $F$ defined from the restriction
$F(c),|c|\le 1$. Taylor-expanding the inverse yields the exact
solution $P_n = \frac{\(nce^{-c}\)^n}{cnn!}$ for $n\ge 1$ for the
asymptotic component size distribution, with the large-$n$ behaviour
$P_n \to \frac{ (ce^{1-c})^n} {\sqrt{2\pi} c n^{3/2}}$, decaying
exponentially for $c\neq 1\Rightarrow ce^{1-c}<1$, but only as a power
for $c=1$.

\subsubsection{Component sizes in IRG}

For an IRG model, the asymptotic component size distribution, and thus
its generating function $g(z)$, will obviously be an average over the
result conditional upon a particular color of the initial node, and we
can write
\beq
\label{g_IRG}
	g(z) = \sum_a r_a g_a(z),
\eeq
where $g_a(z)$ is conditional upon initial color $a$; these satisfy
the following recursive relations.
\beq
\label{gg_IRG}
	g_a(z) = z \exp\left[ \sum_b c_{ab} r_b\(g_b(z)-1\) \right],
\eeq
an obvious generalization of the corresponding RG result,
eq. (\ref{g_RG}). This cannot in general be solved exactly, but can be
analyzed using numerical and/or series expansion methods.

\subsubsection{Component sizes in DRG}

Next we wish to obtain the asymptotic size distribution $\{P_n\}$ in a
DRG model.  As before, the sparsity forces finite components to take
the form of tree.

The generating function $H(x)$ for the degree distribution will turn
out to be convenient.  Of interest is also the degree distribution of
a node reached by following a random edge. This yields a weighting of
nodes by their degree, resulting in the modified distribution $q_m = m
p_m / \ave{m}$.  The generating function for its {\em remaining}
degree (disregarding the incoming stub) becomes $H'(x)/H'(1)$.

\begin{figure}[h]
\centering
\includegraphics[width=100mm]{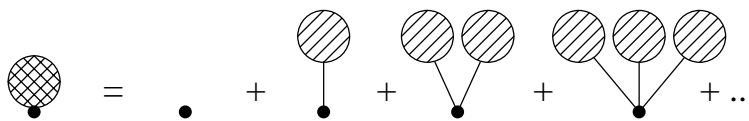}
\caption{$g(z)$ in terms of $h(z)$ for a DRG model, illustrating eq. (\ref{ghDRG}).}
\label{figgh}
\end{figure}

\begin{figure}[h]
\centering
\includegraphics[width=100mm]{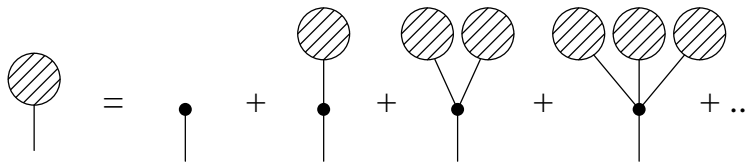}
\caption{Illustration of the recursive relation (\ref{hhDRG}) for $h(z)$.}
\label{fighh}
\end{figure}

With $g(z)$ as before the generating function for the size
distribution of the entire component, let $h(z)$ be the analogous
generating function for the size distribution of a subtree found by
following an edge. Then $g(z)$ can be expressed in terms of $h(z)$ as
\beq
\label{ghDRG}
	g(z) = z \sum_m p_m h(z)^m \equiv z H(h(z)),
\eeq
to be interpreted as follows. The explicit factor of $z$ represents
the first node. It has $m$ outgoing edges with probability $p_m$, each
of which represents a subtree and yields a factor $h(z)$; see
fig. \ref{figgh} for a graphical illustration.
By a similar argument, $h(z)$ satisfies the recursion
\beq
\label{hhDRG}
	h(z) = z \sum_m \frac{m p_m}{\ave{m}} h(z)^{m-1}
	= z \frac{H'(h(z))}{H'(1)},
\eeq
as depicted in fig. \ref{fighh}.
Note that for the case of a Poissonian degree distribution, the
recursion simplifies to the RG result. Indeed, the Poissonian
restriction of DRG is asymptotically equivalent to a version of RG
allowing for non-simple graphs.

\subsubsection{Component sizes in CDRG}

Finally, we wish to obtain the asymptotic size distribution $\{P_n\}$,
and its associated generating function $g(z)$, in a CDRG model.

Here, the multivariate generating function $H(\x)$ for the colored
degree distribution will be needed.  Of interest is also the colored
degree distribution $q_{\m|a}$ of a node reached by following a random
edge emanating from a stub of given color $a$. This is given by
$q_{\m|a} = \sum_b T_{ab} m_b p_{\m}$.  It follows that the generating
function for the distribution of its {\em remaining} colored degree
(where the incoming stub is neglected) is $\sum_b T_{ab}\partial_b
H(\x)$.

With $g(z)$ having its usual meaning, we will denote by $h_a(z)$ the
analogous generating function for the size distribution of a subtree
found by following an edge emanating from a stub of color $a$. Then,
generalizing eq. (\ref{ghDRG}), $g(z)$ can be expressed in terms of
$\h(z)=(h_1(z),\dots,h_K(z))$ as
\beq
\label{gh_CDRG}
	g(z) = z \sum_{\m} p_{\m} \prod_a h_a(z)^{m_a} \equiv z H(\h(z)),
\eeq
with the following interpretation. The explicit factor of $z$ accounts
for the first node, which has a colored degree $\m$ with probability
$p_{\m}$; each stub of color $a$ represents a subtree and yields a
factor $h_a(z)$. The argument generalizes the one used for DRG, as
depicted in fig. \ref{figgh}.

By a similar argument, $\h(z)$ satisfies the coupled recursion
\beq
\label{hh_CDRG}
	h_a(z) = z \sum_b T_{ab} \partial_b H(\h(z),
\eeq
generalizing the DRG relation, eq. (\ref{hhDRG}), depicted in
fig. \ref{fighh}.

\subsection{The Appearence of the Giant}

For all models, $g(z)$ is the generating function for the component
size distribution $\{P_n\}$, and normalization of probability requires
$g(1)=1$.  Indeed, this corresponds to a fixed point of the recursions
for $z=1$ in all models. However, it is a physical solution only if it
corresponds to a stable fixed point of the associated recursive
equations. Where it fails to be stable, a competing solution with
$g(1)<1$ will take over, yielding a {\em probability deficit} of
magnitude $1-g(1)$.

For the asymptotically resulting branching process this can be
interpreted as being due to a finite probability to obtain an infinite
tree. For a finite but large graph, it corresponds to the appearance
of a {\em giant component}, asymptotically containing a finite
fraction $1-g(1)$ of the nodes, and the transition where the naive
fixed point loses stability defines a {\em percolation threshold},
typically of second order. Below the threshold, all components are
small, and above it there is a single giant, while the remaining
components are small.

\subsubsection{The giant in RG}

For $z=1$, the recursion (\ref{g_RG}) for $g=g(1)$ simplifies to $g\to
e^{c(g-1)}$, with a solution satisfying $cge^{-cg}=ce^{-c}$. The
stability of a solution depends on the magnitude of the Jacobian,
given by $ce^{c(g-1)}$, which equals $cg$ when $g$ is a solution.

It has the trivial solution $cg=c$, i.e. $g=1$.  For $c$ smaller than
a critical value, $c=1$, this solution is indeed stable under
iteration of the recursion, with the Jacobian given by $c$.

For $c>1$, it fails to be stable, and so we must look for another
fixed point as the physical solution. Indeed, such a fixed point
exists, as follows from looking at a graph of the function $c\to
ce^{-c}$, which has a unique maximum for $c=1$. Thus, for each $c>1$
there is a dual value $\hat{c}<1$ with the same value of this
function, yielding the stable solution $g(1)=\hat{c}/c < 1$.

Thus, we have established a probability deficit for $c>1$, reflecting
the existence of a giant component, asymptotically containing a finite
fraction $1-\hat{c}/c$ of the nodes. The critical point
$c=1\Rightarrow \hat{c}=1$ defines the percolation threshold, above
which there is a finite probability for an arbitrary pair of nodes to
be connected via a finite path.

\subsubsection{The giant in IRG}

For an IRG model, $g=g(1)$ is given by the linear combination
$g=\r\cdot\g \equiv \sum_ar_ag_a$, with $\g=\g(1)$ satifying the
recursion $g_a \to \exp[\sum_bc_{ab}r_b(g_b-1)]$, as follows from
setting $z=1$ in eqs. (\ref{g_IRG},\ref{gg_IRG}). The stability of the
trivial solution $\g=\1$ depends on the spectrum of the local Jacobian
matrix $\J=\{c_{ab}r_b\}$.

The case of the largest eigenvalue of $\J$ being exactly unity defines
a critical hypersurface in parameter space, beyond which the trivial
fixed point $\g(1)=\1\Rightarrow g(1)=1$ loses stability, and a
competing fixed point appears with $g_a(1)<1\Rightarrow g(1)<
1$. Again, the corresponding probability deficit $1-g(1)$ is taken as
the probability for winding up in a giant component of size
$N(1-g(1))$.

\subsubsection{The giant in DRG}

For a DRG model, setting $z=1$ in eqs. (\ref{ghDRG},\ref{hhDRG}),
yields for $g=g(1)$ and $h=h(1)$ the relation $g=H(h)$ and the
recursion $h\to H'(h)/H'(1)$, with the trivial fixed point
$h=1\Rightarrow g=1$. The stability of this is governed by the
Jacobian $H''(1)/H'(1) = \ave{m(m-1)}/\ave{m}$. Stability results if
this is smaller than unity, i.e. if $\ave{m(m-2)}<0$, defining the
subcritical domain of DRG \cite{MoRe98}.

In the supercritical domain, there will be a unique competing solution
$h<1$, satisfying $hH'(1)=H'(h)$, yielding $g<1$, with the
corresponding probability deficit indicating the existence of a giant
component.

\subsubsection{The giant in CDRG}

Similarly, in a CDRG model, we can pinpoint the subcritical region by
analyzing the stability of the trivial solution $\h(1)=\1$ of the
recursion (\ref{hh_CDRG}) with $z=1$, amounting to $\h\to \T\partial
H(\h)$.  The Jacobian amounts to $\J=\T\E$, i.e. the matrix product of
$\T$ and the matrix $\E=\{E_{ab}\}$ of second order multivariate
combinatorial moments of the colored degree distribution, as defined
in eq. (\ref{MCMom}), and subcriticality corresponds to the largest
eigenvalue of $\J$ being smaller than unity.

In the supercritical region, we will have non-trivial solution
yielding $g(1)<1$, with an associated probability deficit and a giant
component of corresponding relative size.

\section{Discussion}

All of the models discussed in this article admit versions with or
without the restriction to simple graphs. They share the existence of
several nice properties, such as the computability of interesting
local and global characteristics, and the existence of a phase
transition in the form of a percolation threshold, where a giant
component appears.

The sparse RG model is a mathematically very interesting
object. Nevertheless, it is severly limited as a model of real-world
networks. Its degree distribution is restricted to be Poissonian, and
it is suffers from a fundamental lack of correlations between
edges. Its main importance is as a role model for more general random
graph models.

The DRG approach yields a general class of random graph models, and
contains a non-simple version of RG as a special case.  Although it
admits arbitrary degree distributions, it shares with RG a fundamental
lack of edge correlations.

Generalizing RG by adding hidden variables in the form of unobservable
vertex colors, allowed to affect edge probabilities, yields another
general class of models -- IRG. It admits arbitrarily many distinct
models for a single degree distribution, and displays non-trivial edge
correlation. Its most serious limitation lies in the restriction of
the degree distribution to a Poissonian mix (which however does not
exclude power-behaviour!). It trivially contains RG as a special case,
and its restriction to a rank one preference matrix, $c_{ab}=C_aC_b$,
defines a class of uncorrelated models, that has been shown to be
asymptotically equivalent to the restriction of DRG to Poissonian
mixtures \cite{Sod02}. Thus, IRG and DRG define distinct superclasses
of the classic RG model, and one might expect that there exists a
larger class that contains them both as distinct restrictions.

Such a unified class of models indeed exist. The generalization of DRG
to models with unobservable color on individual stubs, that is allowed
to affect the edge probabilities as emerging from the stub pairing
statistics, yields a very general class of models -- CDRG.  It allows
for arbitrary degree distributions, as well as for non-trivial edge
correlations. It contains as distinct subclasses both DRG (trivially)
and IRG (as the restriction of the colored degree distribution to a
mix of multivariate Poissonians) \cite{CDRG,CDRG2}.

CDRG shares with DRG an interesting relation to Feynman graphs of
simple field theories; work is in progress to explore this relation.
CDRG should also admit a straightforward extension to cover models
also of directed graphs.

A unifying formalism for random graphs appears to be a prerequisite
for the possibility to devise a systematic model inference scheme
based on the observed properties of real-world networks.  CDRG appears
to be a step on the way to such a formalism for sparse, truly random
graphs.

\section*{Acknowledgment}

The author thanks the organizers for bringing together a highly
interesting workshop, and for the extra bonus of being able to
experience the beautiful environment offered by central Krakow.

This work was in part supported by the Swedish Foundation for
Strategic Research.



\providecommand{\bysame}{\leavevmode\hbox to3em{\hrulefill}\thinspace}
\providecommand{\MR}{\relax\ifhmode\unskip\space\fi MR }
\providecommand{\MRhref}[2]{%
  \href{http://www.ams.org/mathscinet-getitem?mr=#1}{#2}
}
\providecommand{\href}[2]{#2}

\end{document}